\PassOptionsToPackage{usenames,dvipsnames}{xcolor}
\documentclass{article}

\usepackage{microtype}
\usepackage{graphicx}
\usepackage{booktabs} 

\usepackage[colorlinks=true]{hyperref}
\hypersetup{breaklinks=true}
\usepackage[many]{tcolorbox}
\usepackage{pifont}
\usepackage{dsfont}
\usepackage{hwemoji}

\usepackage[accepted]{icml2025}

\usepackage{amsmath}
\usepackage{amssymb}
\usepackage{mathtools}
\usepackage{amsthm}
\usepackage{enumitem}
\usepackage{setspace}
\usepackage{multirow}
\usepackage{color}
\usepackage{soul}
\usepackage{multicol}
\usepackage[capitalize,noabbrev]{cleveref}
\usepackage[caption=false, font=footnotesize, captionskip=0pt]{subfig}

\theoremstyle{definition}

\newtheorem{exm}{Example}[section]

\theoremstyle{definition}

\theoremstyle{remark}

\usepackage[textsize=tiny]{todonotes}

\icmltitlerunning{Reliable agent engineering should integrate machine-compatible organizational principles}

\begin{document}

\twocolumn[
\icmltitle{Reliable agent engineering should integrate\\ machine-compatible organizational principles}




\begin{icmlauthorlist}
\icmlauthor{R. Patrick Xian}{kho}
\icmlauthor{Garry A. Gabison}{qmul}
\icmlauthor{Ahmed Alaa}{ucb,ucsf}
\icmlauthor{Christoph Riedl}{dkb,kho,nsi}
\icmlauthor{Grigorios G. Chrysos}{wima}
\end{icmlauthorlist}

\icmlaffiliation{kho}{Khoury College of Computer Sciences, Northeastern University, Boston, MA, USA}
\icmlaffiliation{dkb}{D'Amore-McKim School of Business, Northeastern University, Boston, MA, USA}
\icmlaffiliation{nsi}{Network Science Institute, Northeastern University, Boston, MA, USA}
\icmlaffiliation{qmul}{Centre for Commercial Law Studies, Queen Mary University of London, London, UK}
\icmlaffiliation{ucb}{Department of Electrical Engineering and Computer Science, University of California, Berkeley, Berkeley, CA, USA}
\icmlaffiliation{ucsf}{University of California, San Francisco, San Francisco, CA, USA}
\icmlaffiliation{wima}{Department of Electrical and Computer Engineering, University of Wisconsin--Madison, Madison, WI, USA}



\icmlcorrespondingauthor{R. Patrick Xian}{xrpatrick@gmail.com}
\icmlcorrespondingauthor{Grigorios G. Chrysos}{chrysos@wisc.edu}


\vskip 0.3in
]



\printAffiliationsAndNotice{}  

\begin{abstract}

As AI agents built on large language models (LLMs) become increasingly embedded in society, issues of coordination, control, delegation, and accountability are entangled with concerns over their reliability. To design and implement LLM agents around reliable operations, we should consider the task complexity in the application settings and reduce their limitations while striving to minimize agent failures and optimize resource efficiency. High-functioning human organizations have faced similar balancing issues, which led to evidence-based theories that seek to understand their functioning strategies. We examine the parallels between LLM agents and the compatible frameworks in organization science, focusing on what the design, scaling, and management of organizations can inform agentic systems towards improving reliability. We offer three preliminary accounts of organizational principles for AI agent engineering to attain reliability and effectiveness, through balancing agency and capabilities in agent design, resource constraints and performance benefits in agent scaling, and internal and external mechanisms in agent management. Our work extends the growing exchanges between the operational and governance principles of AI systems and social systems to facilitate system integration.
\end{abstract}

\section{Introduction}
\label{submission}

The construction and adoption of AI agents have drawn great interest \citep{tomasev_virtual_2025,qu_comprehensive_2025} because agentic systems can automate workflows and streamline decision-making across specialized domains \citep{yam_machine_2025}. Despite the tremendous enthusiasm in deploying these systems, the public has limited awareness of their potential consequences \citep{mitchell_fully_2025}. Besides, many latent challenges exist in transferring AI agents from simulated environments to the real world (i.e. Doyle's catch) \citep{woods_risks_2016}.
\begin{figure}[htbp]
\vspace{-0.5em}
    \centering
    \includegraphics[width=0.95\linewidth]{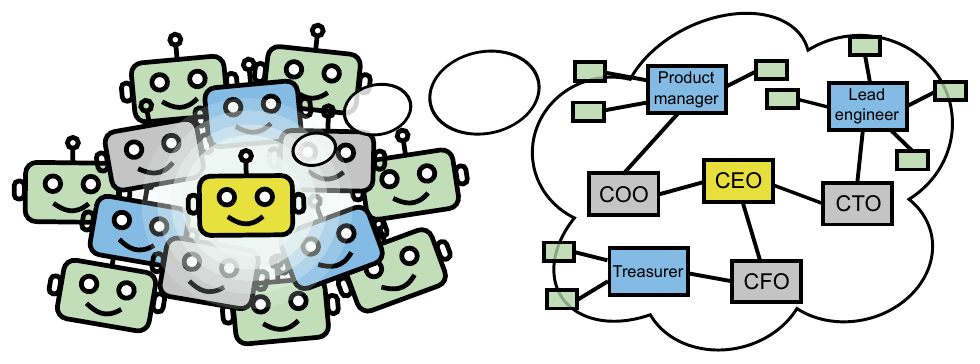}
    \vspace{-0.5em}
    \caption{Machine-compatible organization theory accounts for the issues of AI agents working in groups in contrast to human agents.}
    \vspace{-0.5em}
    \label{fig:agent_orgs}
\end{figure}
At the moment, the increasing empirical evidence from behavioral studies \citep{hagendorff_machine_2023,chang_lmbehav_2024} of LLMs and AI agents lends themselves to formulating a machine-compatible organizational theory (Fig. \ref{fig:agent_orgs}) that can assist agentic system design and quality control, and facilitate their societal integration by leveraging the numerous parallels while accounting for apparent discrepancies with structured human counterparts (Table \ref{tab:orgvsags}). NVIDIA's Jensen Huang even remarked \citep{morse_nvidias_2025} in a 2025 keynote that ``the [Information Technology] department of every company is going to be the [Human Resources] department of AI agents in the future.''
\begin{table*}[htbp!]
\setstretch{1.1}
    \centering
    \begin{tabular}{ccc}
         \toprule
         \textbf{Characteristic} & \textbf{Organization (for-profit)} & \textbf{Agentic system (LLM-based)} \\
         \midrule
         \multirow{2}{*}{Structure} & \multirow{2}{*}{\parbox{5cm}{\setstretch{1}\centering Determined by strategy, size, environment, interactions, etc.}} & \multirow{2}{*}{\parbox{5cm}{\setstretch{1}\centering Determined by function or tasks, engineering constraints, etc.}} \\
         & & \\
         \midrule
         Actors & Bounded rational agents & LLM agents \\
         \midrule
         Goals & Determined by management & Determined by user or provider \\
         \midrule
         \multirow{2}{*}{Incentives} & \multirow{2}{*}{\parbox{5.5cm}{\setstretch{1}\centering Determined by contract, task visibility, risk-reward balance \citep{gibbons_incentives_1998}}} & \multirow{2}{*}{\parbox{5.8cm}{\setstretch{1}\centering Determined by reward function specified in system design or in deployment}} \\
         & & \\
         \midrule
         \multirow{2}{*}{Oversight} & \multirow{2}{*}{\parbox{4.6cm}{\setstretch{1}\centering Determined by directors or board of trustees}} & \multirow{2}{*}{\parbox{5cm}{\setstretch{1}\centering Determined by agent platform creators according to regulations}}\\
         & & \\
         \midrule
         \multirow{2}{*}{Scaling} & \multirow{2}{*}{\parbox{4.6cm}{\setstretch{1}\centering Determined by industry, structure, technology, etc.}} & \multirow{2}{*}{\parbox{4.6cm}{\setstretch{1}\centering Determined by structure, interactions, resources, etc.}}\\
         & & \\
         \bottomrule
    \end{tabular}
    \caption{\setstretch{1}Comparison between standard for-profit organizations and LLM agents in multiple dimensions.}
    \label{tab:orgvsags}
\vspace{-1em}
\end{table*}

Most existing and imminent AI agent applications in the real world target specialized domains (e.g., solving a business task or automating a disjoint, labor-intensive workflow in healthcare and financial sectors). To deliver value, a high success rate is required in task completion while safety and security risks are minimized. These goals are still far from attainable \citep{pan_why_2025,li_commercial_2025,zou_security_2025}, despite the promise of the agentic economy \citep{rothschild_agentic_2025,tomasev_virtual_2025}. Another major drawback for AI agents is that they are considerably more prone to manipulation than monolithic foundation models because of their decentralized nature and sensitivity to defective components \citep{chiang_why_2025}. Given that the reliability bottleneck continues to be a critical issue \citep{yao_taubench_2025,mazeika_remote_2025}, we argue that reliable engineering of AI agents benefits from the elaboration of effective organizational principles compatible with their behavioral traits and capabilities. Besides, understanding these principles can also inform the governance and coordination of AI agents and form a necessary step towards transparent and accountable system integration into human organizations \citep{mathur_advancing_2024,collins_building_2024}.
\vspace{-1em}

\paragraph{Why agentic systems need organizational principles?} Organization science distills principles from human organizations to explain how organizations emerge, evolve, achieve efficiency and productivity, and maintain resilience. Modern organization theory emerged through the studies of industrial practices \citep{march_organizations_1993} but has since been generalized to other types of social, institutional, and digital systems \citep{haveman_power_2022}. \textbf{An organizational (or organizing) principle is any procedure that coordinates work and processes information within or between organizations} \citep{mcevily_trust_2003}. The principles can be determined through internal and external factors associated with the organization. They are a convenient way to rationalize and compare system design and operations \citep{horling_survey_2004}. Classic works in organization science have examined the organization structure \citep{dalton_organization_1980} and organizational behavior of the actors in decision-making \citep{simon_administrative_1997}. Examples of human organizational principles include productivity (e.g. through division of labor) \citep{becker_division_1992}, resource efficiency \citep{hillman_resource_2009}, adaptability \citep{cyert_behavioral_1992}, trust \citep{mcevily_trust_2003}, synergy \citep{dessein_synergy_2010}, resilience \citep{weick_managing_2015}, etc. Similarly, in an agentic system $S$ with many components (e.g. architecture, communication protocol, policy, learning objective, reward signals, etc.), organizational principles can be built into single or multiple components. We regard an organizational principle $H$ as constructive if it improves system performance measured by metric $R$ than a baseline principle $H_0$ for a set of tasks $\mathcal{T}$,
\begin{equation}
    \mathds{E}_{\mathcal{T}}[R(S;H)] > \mathds{E}_{\mathcal{T}}[R(S;H_0)].
    \label{eq:org_eval}
\end{equation}
They are implemented in the design of the organization structure and the behavior of the actors within it.
\vspace{-0.5em}
\begin{figure*}[htb!]
    \centering
    \subfloat[]{
    \includegraphics[width=0.78\linewidth]{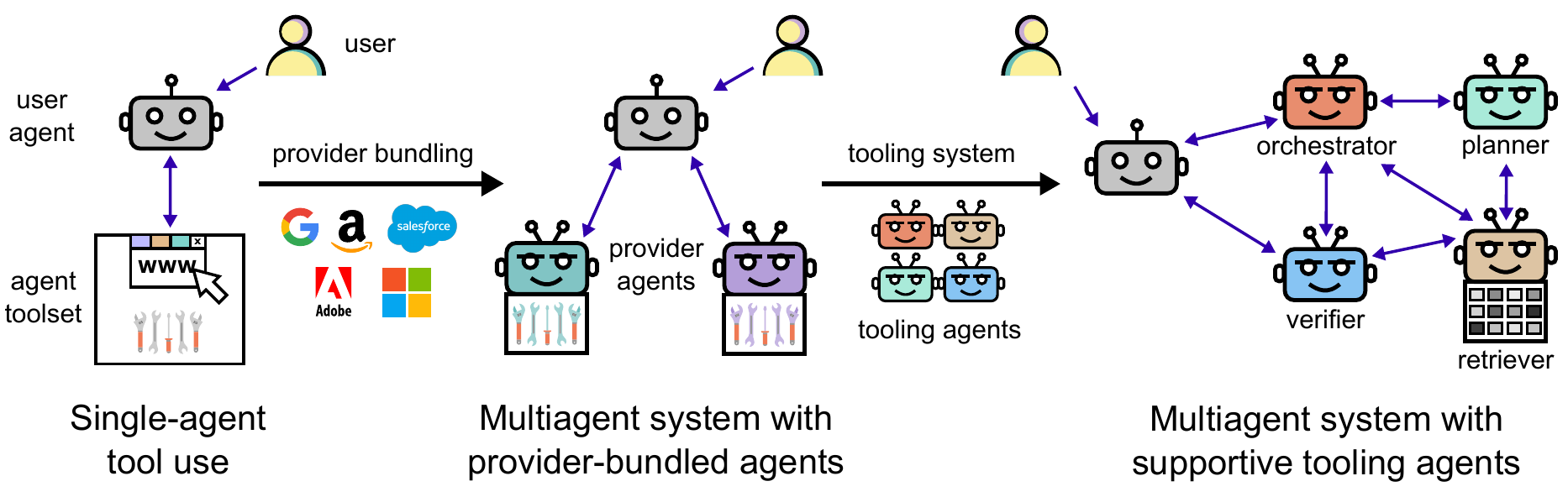}
    }
    \hfill
    \subfloat[]{
    \includegraphics[width=0.16\linewidth]{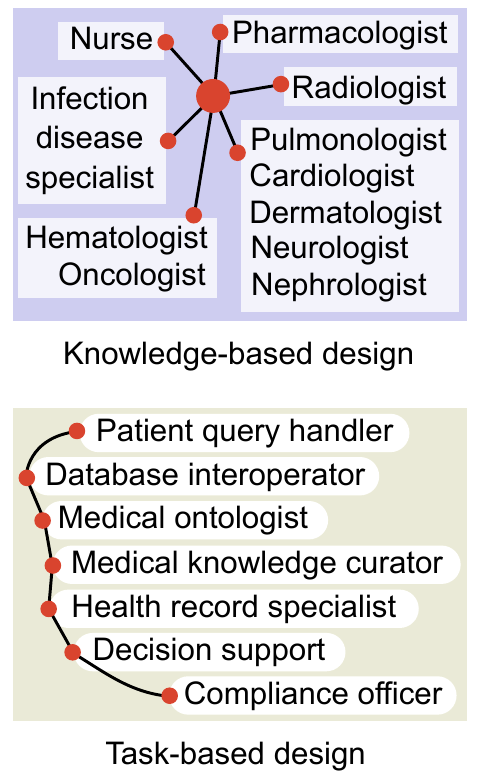}
    }
    \vspace{-0.4em}
    \caption{(a) Tool-use agentic system with distinct delegational structures to handle user requests. Single-agent tool-use has high requirements on user agent capabilities; Provider bundling flows down the requirements externally; Supportive tooling agents can further outsource tool handling to tool providers in exchange for increased system capability. (b) Knowledge-based (as a hierarchical structure) and task-based (as a horizontal workflow) views of organization design for medical agentic systems for patient diagnostics.}
    \label{fig:agent_tooling}
    \vspace{-1em}
\end{figure*}

\paragraph{Why should organizational principles for LLM agents be machine-compatible?} LLM agents are disembodied entities \citep{shanahan_palatable_2025} and their developmental history doesn't contain genuine human experiences nor is it aligned with evolutionary trajectories that shaped human cognition and social behavior \citep{cuskley_limitations_2024}. This creates problems in ascribing them human-like agency in a virtual organization. Although LLM agents can simulate more dimensions of human interactions than software agents \citep{sycara_multiagent_1998,horling_survey_2004,dignum_multiagent_2013} because of their reasoning capabilities, synthetic personalities \citep{serapio-garcia_personality_2023}, and strategic behavior \citep{jia_strategic_2025}, an LLM-based MAS is still fundamentally different from a human organization. \textbf{To properly consider LLM agents from an organizational viewpoint, the operational constraints of their disembodied and anthropomorphic nature should be fully accounted for.} Designers must consider the failure modes of the models and those for the system \citep{pan_why_2025,gabison_inherent_2025}. At the model level, they should consider the limitations in the model development processes \citep{kaddour_challenges_2023,casper_open_2023}. At the system level, they should also account for caveats in agent-tool and agent-agent interactions, the fidelity of agent components, and the discrepancies of agent decision-making from humans \citep{gao_take_2025}.

Next, we illustrate how organizational views can help understand and improve AI agents from distinct angles. Our contributions differ from \citet{miehling_agentic_2025}, which considered a systems-theoretic approach for AI agents but does not provide concrete discussions of LLM-based MAS and the allocation of machine agency to user requests. \citet{kolbjornsrud_designing_2024,collins_building_2024,riedl_ai_2025} discussed various scenarios of integrating AI systems with human teams to improve human performance, but they do not consider the direct impact of AI agent design on reliability. Our viewpoint zooms in on the agentic system, where the agent architecture and interaction patterns are made to serve the human user as part of the environment. The user interacts with the agentic system's output (e.g. AI assistants) or receives the output as a result (e.g. AI agents planning and executing a task) but has partial or indirect observability of individual agent's behavior \citep{yao_taubench_2025}.

\section{Organization design for AI agents}

\subsection{Organization structure and its use}
Organization structures establish the roles, responsibilities, and relationships that enable enterprises to accomplish their objectives \citep{harris_organization_2002,burton_organizational_2020}. The structure affects how information is processed and decisions are made \citep{csaszar_efficient_2012}, determines the efficiency, performance, and resilience of the organization \citep{dalton_organization_1980,joseph_organizational_2020}. A good organization design strives to achieve both internal fit and external fit\footnote{Internal fit means each component can benefit from interacting with others within the structure. External fit requires components to satisfy the information processing need with the environment} \citep{burton_organizational_2020,joseph_organization_2025}. Human organization's operations and development are influenced by the collective forces (e.g. social, legal), conditions (e.g. political, cultural), and factors (e.g. economic, ecological), referred to as the organizational environment \citep{aldrich_organizations_1979,dess_dimensions_1984}. An effective organization continuously monitors and adapts to changes in its environment, creating strategic responses that maintain alignment between internal capabilities and external demands \citep{cyert_behavioral_1992}.

Besides the critical role of organization structures in facilitating routine operations, they offer an effective way to implement controls (i.e. structure control) on actors within the organization \citep{ouchi_relationship_1977}. This includes the control of task allocation and hierarchy \citep{hart_design_2005}, which affects the agency of the actors, specialization, and their partitioning of shared resources \citep{carroll_organizational_2002}. Structure-level controls can also come from policy requirements of information sharing, such as regulations and security measures, which require well-coordinated information flow.

\subsection{Organizational view of AI agent design}
An agent architecture is the basis for allocating and coordinating tasks \citep{campbell_mas_2011}. A central question in designing agentic systems is whether the performance gain from multiple specialized agents outweighs the increased failure modes due to miscoordination \citep{sycara_multiagent_1998}. In organizational settings, this leads to \textit{structural differentiation} \citep{blau_formal_1970}, which proposed solutions according to specific contexts. One primary concern for AI agents is that single agents with highly centralized and concentrated intelligence can exacerbate safety and misalignment concerns\footnote{``Centralized'' indicates that the system is governed by a small group of agents; ``Concentrated'' indicates that (complex) actions are initialized and performed by a small group of agents.} \citep{mitchell_fully_2025,greenblatt_why_2025}. These agents exhibit flawed agency and unconventional failure patterns \citep{gabison_inherent_2025} in comparison with human agents, which create serious hurdles when they have access to sensitive information and resources \citep{john_owasp_2025,zou_security_2025}. Moreover, design decisions made by agent providers determine engineering accountability and legal liability in the event of system failures that are often less appreciated initially than performance. Therefore, answering the agent design question also requires weighing factors beyond technical capabilities, such as risk exposure and operational priorities, as illustrated below.

\begin{exm}[Tool-use agentic systems]\label{exm:tool_agents} Tool-use by LLM agents \citep{wang_what_2024} is a common way to expand agent capability and usability (Fig. \ref{fig:agent_tooling}a). The user tasks are delegated to a user agent, which seeks the relevant tools (e.g. APIs, software) to complete the task. However, this setting also created many potential risks \citep{ye_toolsword_2024}. Alongside engineering constraints \citep{zhou_masdesign_2025}, agentic system design should strategically consider architectures (Fig. \ref{fig:agent_tooling}a) with a variety of delegational structures \citep{castelfranchi_delegation_1998} that balance user task complexity with system requirements and risks through effective outsourcing \citep{aron_just_2005} to achieve reliability as discussed below and in Appendix \ref{app:tools}.
\begin{enumerate}[wide, labelindent=0pt, itemsep=0pt, topsep=1pt]
    \item[\textbullet] \textbf{Single-agent tool use} has the most straightforward agent architecture \citep{wang_what_2024}. The user agent selects the necessary tool to execute a task from the available toolset. The tool selection requires reasoning, but can also be guided with human instructions. It requires the user agent to understand and compare the functionalities of all available tools. Therefore, the user agent needs to come from a highly capable base model, which increases the chance of misaligned behavior \citep{greenblatt_why_2025}. Access to a large number of tools can also increase the chance of tool hallucination \citep{patil_gorilla_2024,langchain_benchmarking_2025}.
    \item[\textbullet] \textbf{MAS with provider-bundled agents} distributes the tool selection to external software providers, which then deploy tooling agents and with which the user agent interacts. The provider bundles services together separately for marketing purposes \citep{gandal_aint_2018} to strategically minimize information sharing. The user agent does not need the capability to parse through and understand the full content of external software, but only to communicate with provider-bundled tooling agents on tool selection and calling procedures through interactions with them. This simplifies the requirements in base model quality for the user agent.
    \item[\textbullet] \textbf{MAS with supportive tooling agents} distributes the tool access to individual agents specialized in distinct steps of an agent tooling service.\footnote{Tooling services are now a part of the agentic market and the setting considered here is a simplified version.} The action sequence (i.e. agent trajectory) starts from the user agent communicating with the orchestrator, which then allocates the tasks to other worker agents (e.g. planner, retriever, verifier) \citep{shi_learning_2024,lu_octotools_2025}. This architecture further distributes the agency among different supportive agents that could be outsourced to other providers. It enables the tuning of each tooling agent individually to carry out the assigned task through coordination between heterogeneous agents\footnote{Heterogeneous agents come from different providers, which may have complementary skills and distinct safety standards.}.
\end{enumerate}
\end{exm}
\begin{exm}[Medical agentic systems] Medical decision-making in the real world is a prototypical example of iterative reasoning under uncertainty \citep{sox_medical_2024}. The process also involves information aggregation from different medical professions, a characteristic phenomenon in organizational decision-making \citep{csaszar_organizational_2013,joseph_organizational_2020}. Current designs of medical AI agents generally take two major approaches representative of other specialized LLM agents (Fig. \ref{fig:agent_tooling}b): The first is to partition work into different knowledge domains \citep{kogut_knowledge_1992} using specialized agents that mimic the professional distinctions within the medical faculties \citet{tang_medagents_2024,kim_mdagents_2024} and aggregates opinions from the specialized agents for decision-making. The other approach is to partition work into tasks and assign a specialized agent in charge of a small number of tasks in a standardized workflow \citep{jiang_medagentbench_2025}. These two design approaches have their respective characteristics: The knowledge-based design can achieve effective selection of experts in adaptive decision-making \citep{kim_mdagents_2024} and enables the use of multiagent debate for performance gain. The task-based design allows a convenient layout for sequential debugging and enforcing compliance because decision-making proceeds in a predictable path. More elaborate medical agentic systems have combined these two design approaches \citep{tu_towards_2025,palepu_towards_2025} to form cross-functional teams. However, standard theories of organization design do not yet account for the fundamental problems in LLM agents, such as the limited consistency of agent interactions within their assigned roles in multi-turn settings \citep{laban_lost_2025} and the validity of performance evaluation \citep{alaa_position_2025} for optimizing or selecting specific designs.



\end{exm}
\begin{tcolorbox}[colback=Gray!20, colframe=Black, left=.2mm, right=.2mm, top=.5mm, bottom=.5mm, sharp corners, boxrule=-1pt, label=rmk1, segmentation style=solid]
\textbf{Remark 1} Machine-compatible agent design should consider the balance of capabilities among individual agents such that the system maintains consistency in repeated decision-making and in performance evaluation.
\end{tcolorbox}

\section{Organizational scaling for AI agents}

\subsection{Economies (and diseconomies) of scaling}
Organizations often scale up their operation when growing, which introduces tradeoffs between efficiency gains and coordination costs. The theories of firm growth highlight the benefit of firms from increasing returns to scale through specialization, resource pooling, and cost distribution. These represent \textit{economies of scale} \citep{stigler_economies_1958}, leading to reduced per-unit production costs: as the input increases, the output increases faster (Fig. \ref{fig:agent_scale}a). Scaling allows units within a company to specialize, invest in research and development \citep{symeonidis1996innovation}, or open up new lines of business \citep{hall1967firm}. As firms expand, they can also benefit from increasing the tasks they perform to manufacture multiple products, which is referred to as \textit{economies of scope} \citep{panzar_economies_1981}. It leads to efficiency improvement linked to the reduction of per-unit production costs for tasks carried out together rather than separately or by combining tools.

The opposites of the growth trends represent diseconomies. Diseconomies of scale (Fig. \ref{fig:agent_scale}a) can occur because some resources (e.g., good managers, unique talents) cannot be replicated. Another reason is that communication pathways within the organization multiply exponentially, which can slow down processes and increase bureaucratic overhead \citep{amato1985effects}. Diseconomies of scope occur when a company becomes less efficient by expanding into too many different products or services, leading to the costs of managing diverse business activities outweighing the benefits of shared resources and capabilities \citep{bresnahan_schumpeterian_2012}. The analysis of organizational scaling in firms often comes down to identifying tradeoff points between economies and diseconomies. Besides, as the firm grows, managing cost (including monitoring and coordination) \citep{robinson_problem_1934} and conflict management \citep{rahim_measurement_1983} are also increasing points of consideration.

\subsection{Organizational view of AI agent scaling}
Scaling laws in deep learning show that relationships between model size, dataset dimensions, and performance capabilities exist across architecture types \citep{kaplan_scaling_2020,li_misfitting_2025}. The inverse scaling effects for certain capabilities observed in LLMs \citep{mckenzie_inverse_2023,wei_inverse_2023} indicate that simply making the model components larger may not always lead to better outcomes. Reliability and effectiveness are two key components in responsible AI scaling \citep{anthropic_responsible_2023,ibm_scaling_2025}. They ensure that the increased cost in scaling from resource consumption and occurrences of misalignment is offset by sufficient performance gain, while the negative side effects are sufficiently controlled. AI agents are software modularized into human-like roles. They gain performance through different forms of scaffolding\footnote{\url{https://aisafety.info/questions/NM25/What-is-scaffolding}} \citep{rosser_agentbreeder_2025} in addition to adopting more powerful base models. Their composite nature indicates that resource consumption, maintenance, and operational compatibility\footnote{Operational compatibility considers how a new or updated framework (here referring to a scaled-up version) is relatable to the procedure of an existing task \citep{karahanna_compat_2006}.} should also be realistically accounted for in system-level scaling besides performance gain. The construct of agents presents opportunities to consider separate scaling regimes (Table \ref{tab:scaling}) \citep{chen_are_2024} in addition to scaling monolithic LLMs.
\begin{table}[htbp!]
\begin{tcolorbox}[colback=NavyBlue!10, colframe=Black, left=.5mm, right=.5mm, top=.5mm, bottom=1mm, rounded corners, label=concepts, segmentation style=solid, equal height group=bottombox]
\setlength\columnsep{1em}
\normalsize
\setstretch{0.9}
\textbf{Structure scaling} considers task execution of an increasing number of agents either through parallel operations and ensembling \citep{li_more_2024} or through increasing the complexity of organization structure by diversifying agent roles (Fig. \ref{fig:agent_tooling}) or modularizing multiagent units \citep{dessein_synergy_2010} to handle complex tasks.\vspace{-5pt}
\tcbline
\vspace{-5pt}
\textbf{Interaction scaling} creates more types of interaction patterns in an agentic system by adjusting communication protocols. Examples include having more rounds of interactions among agents to improve existing behavior using test-time computation \citep{wu_inference_2025} and multiagent collaboration \citep{tran_macollab_2025}.\vspace{-5pt}
\tcbline
\vspace{-5pt}
\textbf{Resource scaling} allows the agents to improve existing resources, such as memory \citep{ouyang_reasoningbank_2025} and environment \citep{camelai_scaling_2025,andrews_are_2025}. Increasing memory allows agents to record a longer sequence of actions and digest a greater amount of information. Increasing the diversity of environments allows agents to be trained to adapt to different contexts.\vspace{-5pt}
\tcbline
\vspace{-5pt}
\textbf{Capability scaling} allows the agents to acquire new capabilities. It can be achieved by changing the base model of agents \citep{belcak_small_2025} or increasing the number of tools \citep{patil_gorilla_2024,gao_democratizing_2025} that agents can access. It can also come from training the agents to learn the combinatorial use of existing tools.
\end{tcolorbox}
\vspace{-1em}
\caption{Different aspects of agent scaling and their manifestations.}
\vspace{-1em}
\label{tab:scaling}
\end{table}

On this note, studies on the growth patterns of organizations (especially firms) provide recipes for understanding agent scaling. For example, capability scaling may allow the system to achieve economies of scope by offering a growing range of agentic services. Resource scaling can lead to an economy of scale through parallel exploration by spawning more agents with shared long-term memory \citep{rezazadeh_collaborative_2025}, which has no exact equivalent in human organizations. Tradeoffs in agent scaling may be quantified using \textit{cost-benefit analysis} of different designs \citep{mishan_costbenefit_2020} by examining how resource-normalized performance changes as the agentic system expands. Such analysis can compare largely distinct designs (e.g. hierarchical vs. horizontal as in Fig. \ref{fig:agent_orgs}b) or designs with granular distinctions at the component level \citep{ziv2000information}. For example, comparing two designs differing by one agent \citep{li_more_2024} only needs to consider the marginal impact of the extra agent. To achieve cost-beneficial scaling, one should focus on the scaling regime and design hyperparameters \citep{wang_efficient_2025} that lead to a significant difference between cost and benefit. Identifying these factors in agentic systems can benefit from examining the performance dependencies from an organizational view, which is illustrated below.

\begin{figure*}[htbp!]
\vspace{-1em}
    \centering
    \subfloat[]{
    \includegraphics[width=0.33\textwidth]{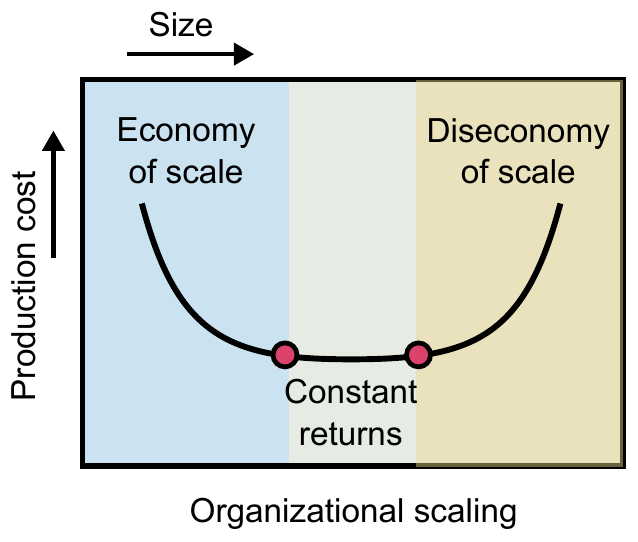}
    }
    \hspace{2em}
    \subfloat[]{
    \includegraphics[width=0.58\textwidth]{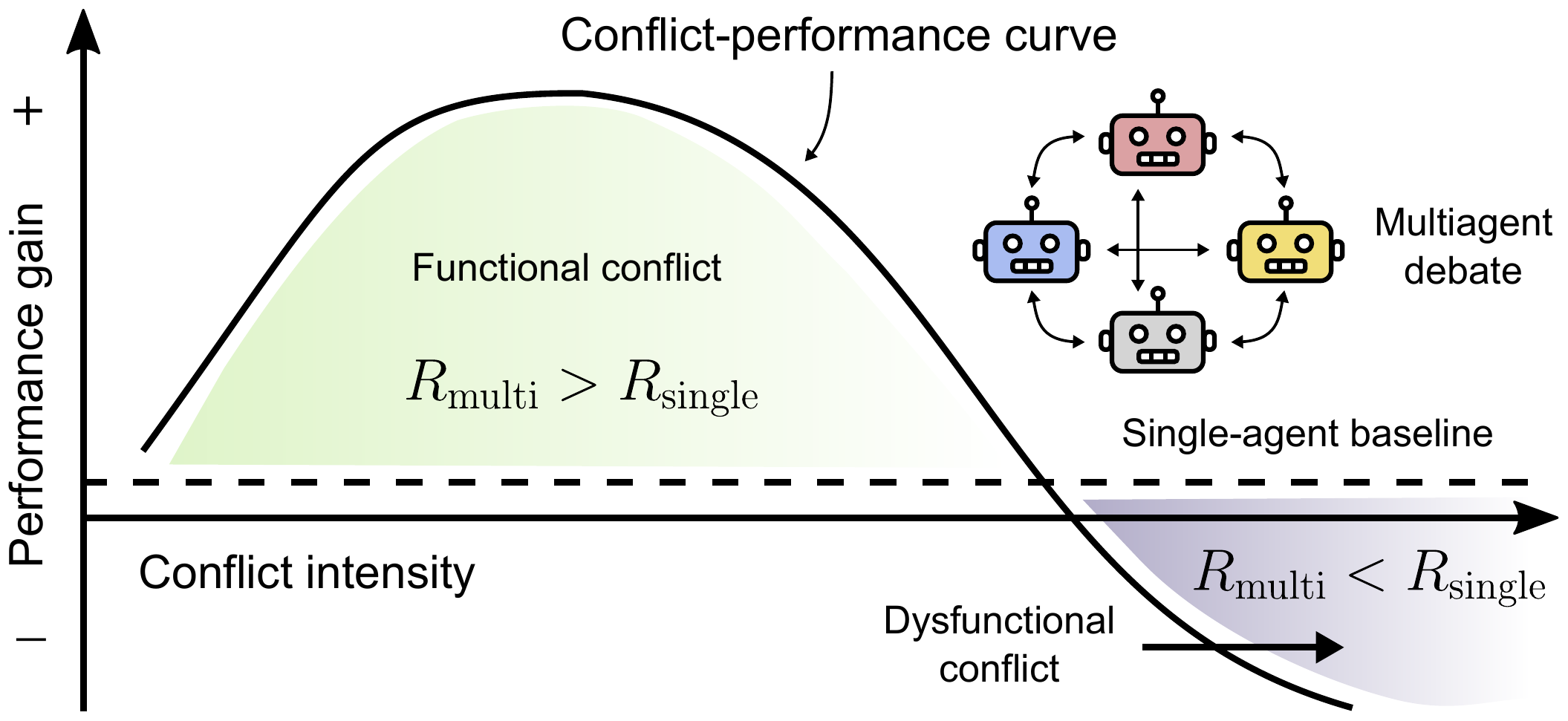}
    }
    \caption{(a) Illustration of the economies and diseconomies in the scaling of firms (i.e. for-profit organizations). (b) Illustration of multiagent debate viewed as organizational conflict within a group which exhibits functional and dysfunctional zones ($R$ is a performance metric as evaluated in Eq. \eqref{eq:org_eval}). ``Conflict intensity'' is controllable by the number of agents, number of turns in a debate, etc.}
    \label{fig:agent_scale}
    \vspace{-1em}
\end{figure*}
\begin{exm}[Maintenance cost in scaling] Agentic systems, like other AI systems, are subject to maintenance due to real-world distribution shifts and performance degradation \citep{chen_airob_2023}. Two main aspects of software maintenance are monitoring and adaptation \citep{banker_software_1993,banker_field_1997,feng_clinical_2022}. They inform the system design and scaling, and contribute differently to maintenance. Monitoring is the \textit{passive} component that can occur continuously or on a recurring schedule around the time of software updates. Adaptation is the \textit{active} component that requires modifying (e.g. model finetuning, updating reasoning or exploration strategies) or augmenting (e.g. addition and expansion of components) parts of an existing system. In organizations, the cost of monitoring grows along with their size, but bigger organizations provide more slack resources to be used in adapting to a changing environment \citep{cyert_behavioral_1992,josefy_all_2015}. From AI agents, the cost of monitoring agent behaviors lies in the detection of erroneous reasoning and planning \citep{korbak_cot_2025,emmons_pragmatic_2025}, along with covert actions that can lead to specification gaming \citep{bondarenko_demonstrating_2025}. Likewise, the monitoring cost also grows as agentic systems scale. For example, monitoring general-purpose agentic systems requires a great amount of resources due to their very large action space compared with domain-specialized systems \citep{hendrycks_unsolved_2021}, they can be adapted easily to different tasks. Domain-specialized AI agents, such as those built on specialized base models, are easier to monitor because of the restricted task space they operate in and the short interaction horizon allowed for task execution \citep{kumar_aligned_2025}. Nevertheless, their cost of adaptation to running additional tasks or dataset shifts is higher because updating the workflow would require retuning base models while balancing task performance. Ultimately, the rate of adaptation required for AI agents in a specific domain hinges on how fast the corresponding environment changes in reality \citep{zheng_gpt4v_2024,lu_build_2025}. As an example, web agents experiencing website updates will need more frequent changes than medical agents experiencing population shifts in patient record databases.
\end{exm}
\begin{exm}[Multiagent debate] As a test-time scaling method \citep{zhang_tts_2025} for improving system performance, multiagent debate \citep{irving_ai_2018} involves agents taking turns to generate responses to a task prompt until a consensus is achieved. In organization settings, this scenario is a form of \textit{intragroup conflict} \citep{jehn_intragroup_2003}, which can be functional or dysfunctional based on the relation between its intensity and performance change \citep{pondy_conflict_1967,amason_distinguishing_1996}. The functional conflict occupies a zone that positively influences organizational performance \citep{pelled_demographic_1996,robbins_essentials_2022}. Multiagent debate with AI agents has been shown to exhibit a similar conflict-performance relationship (Fig. \ref{fig:agent_scale}b), including a nuanced dependence on the agent base models and topics \citep{wynn_talk_2025,zhang_stop_2025}. Although the engineering of multiagent debate allows tracing of the consensus-building process from distinct opinions \citep{schweiger_group_1986}, when not orchestrated in moderation, it can lead to performance degradation and echo chamber effects \citep{estornell_debate_2024,wynn_talk_2025,ma_hunger_2025}. The act of debating benefits from the presentation of diverse viewpoints \citep{pelled_demographic_1996}, which, in agentic systems, is the placement of distinct concepts within the context window as proven to be a source of effectiveness \citep{estornell_debate_2024}. To reduce inference scaling, the debate can become more efficient through pruning procedures \citep{estornell_debate_2024} or employing relaxed turn-taking protocols\footnote{A relaxed turn-taking communication protocol need not involve all agents acting in a fixed order in every turn.} by optimizing the communication topology \citep{li_improving_2024}. These ensure that a functional conflict is reached more easily and consistently and the scaling advantage is reliable.
\end{exm}
\begin{tcolorbox}[colback=Gray!20, colframe=Black, left=.2mm, right=.2mm, top=.5mm, bottom=.5mm, sharp corners, boxrule=-1pt, label=rmk2, segmentation style=solid]
\textbf{Remark 2} Machine-compatible agent scaling requires sensibly weighing the tradeoff between performance benefits and the engineering overheads in monitoring additional agents and system components.
\end{tcolorbox}


\section{Organizational management for AI agents}

\subsection{Managing organizational changes}
Human organizations face challenges in ensuring the consistency of work performance of the actors and the changing environment, which creates unforeseen situations (i.e. contingencies) \citep{donaldson_contingency_2001}. An organization must manage these challenges either via pre-determined contingency plans or by adapting to evolving circumstances on the fly. Managing these situations requires a dynamic component in the organization structure and the cultivation of adaptive organizational behaviors of individuals \citep{eisenhardt_dynamic_2000}. Besides improving technology and workers' knowledge, organizations also invest in shaping worker motivation through management strategies \citep{rainey_work_2000,kanfer_motivation_2016}. In organizational behavior, understanding the relations between motivation and performance is a basis for implementing behavior modification approaches, such as through the scheduling of reinforcement and punishment signals (e.g. materials or financial means), and the setting of tangible goals \citep{locke_theory_1990}. 





Thorough transformations of organizations toward higher productivity or reliability can be achieved by facilitating positive organizational changes \citep{barnett_modeling_1995,weick_organizational_1999}. Analysis of the structure and functioning of safety-critical organizations shows that organizations operating in complex, high-risk environments yet constantly maintaining exceptional safety and performance records follow a set of high reliability (HR) principles \citep{weick_managing_2015,roberts_highrel_2015}: (i) Preoccupation with failure to address potential problems. (ii) Reluctance to simplify unless they are warranted. (iii) Sensitivity to operations by maintaining situational awareness of frontline activities rather than through experience only. (iv) Commitment to resilience, which requires developing capabilities to detect, contain, and recover from errors. (v) Deference to expertise by respecting knowledge over hierarchy during critical situations. These principles have guided practical implementations in reducing failures and errors within human organizations \citep{roe_hrm_2008,casler_revisiting_2014}.

\subsection{Organizational view of AI agent management}
System reliability \citep{pan_why_2025} and security threats \citep{witt_open_2025,zou_security_2025} are among the most critical issues of LLM agents. These issues can multiply during system operation because of the externalities at the user level. Similar to managing organizations to reduce errors, managing LLM agents should therefore consider the coordination across different failure prevention procedures. The behavior of LLM agents can be shaped through mechanisms internal and external to the system to facilitate an organizational change \citep{weick_organizational_1999}.
\begin{figure}[htb!]
    \centering
    \subfloat[]{
    \includegraphics[width=0.85\linewidth]{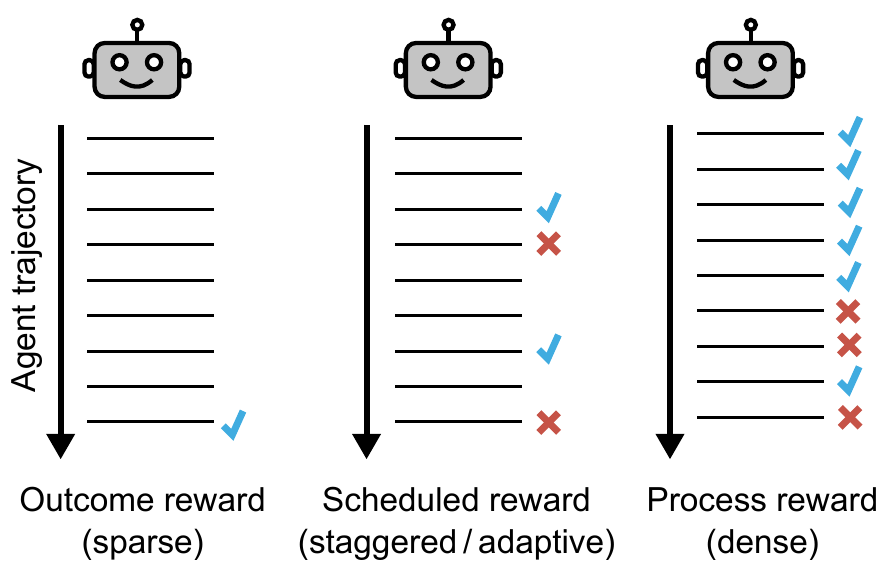}
    }
    \hfill
    \subfloat[]{
    \includegraphics[width=0.49\textwidth]{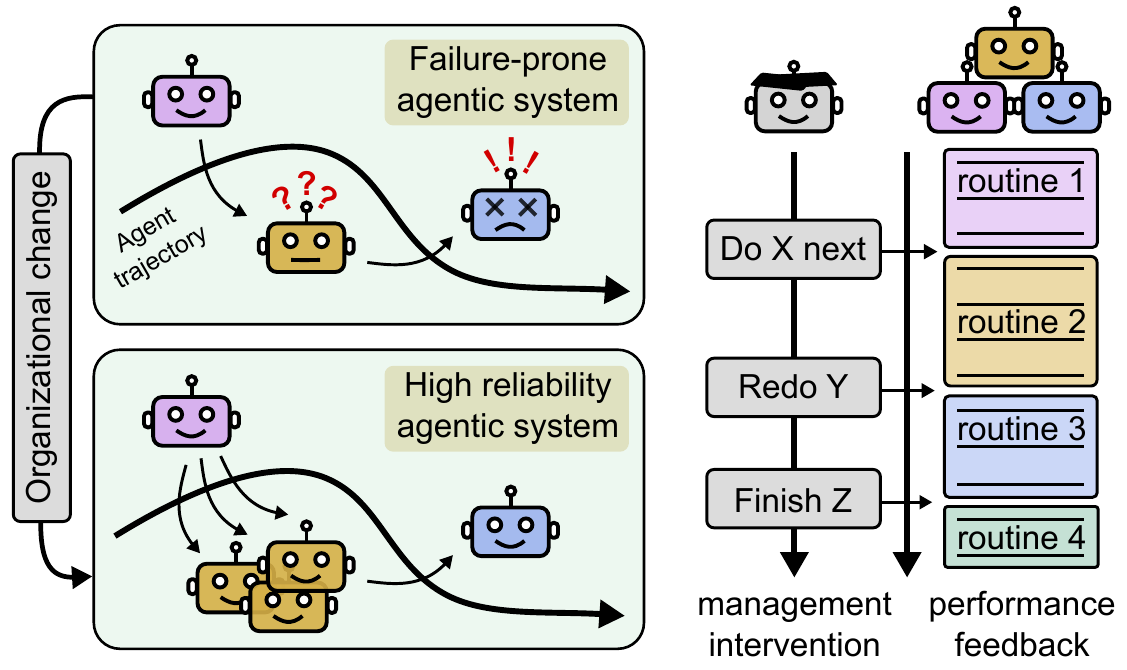}
    }
    \caption{(a) Design of agent reward according to different types of reward feedback (indicated as \ding{56} and \ding{52}) in the trajectory of an agent action sequence (horizontal bars). (b) Improving the reliability of AI agents by cultivating an organizational change (left) and a potential implementation (right) with management intervention and performance feedback.}
    \label{fig:agent_reward}
    \vspace{-1.5em}
\end{figure}

For managing AI agents, \textbf{internal mechanisms} consider the agent configuration and agent interactions within the agentic systems for behavior management. They require little additional resources but depend on the initial settings of the agentic system (e.g. base model, decoding hyperparameters), which are not always accessible to end users. Implementing internal mechanisms can also rely on adjusting role-playing prompts \citep{chen_persona_2024} during agent configuration, which effectively changes the inductive bias of individual agents. Another important aspect is agent interactions, which contain self-reflective behaviors at the single-agent level, including metacognition \citep{didolkar_metacognitive_2024}, self-critique \citep{gou_critic_2024}, and self-improving \citep{acikgoz_selfimprov_2025} capabilities. Inter-agent interactions involve multiple agents such as dialogue, debate, and negotiation \citep{abdelnabi_cooperation_2024}. \textbf{External mechanisms} require additional operations (e.g. finetuning, steering, verification) powered by external resources (e.g. human feedback, computing budget, internet restrictions), but can be more flexible in the type of signals received than internal mechanisms. Human users situated in the environment can also directly influence system behavior. For example, verbal tipping \citep{salinas_butterfly_2024}, coercion \citep{geiping_coercing_2024}, and nudging \citep{cherep_llmnudge_2025} have been demonstrated as potential behavioral influencers for LLMs, which can likewise work for agents in corresponding ways. Other parallels in behavior modification are further discussed below with examples.

    
\begin{exm}[Agent reward design] Reward design for AI agents is conceptually equivalent to determining the right signal to tune worker motivation to induce behavior change \citep{staddon_operant_2003,fishbach_structure_2022}, which is a basis for improving organizational performance. The prevalent approach through scheduling of reinforcement shows that behaviors are best adjusted by assigning reward (or punishment) signals at intervals and staggered throughout task execution. Current reward models in turn-based agent interactions primarily explored outcome-based and process-based rewards \citep{zheng_prm_2025} (Fig. \ref{fig:agent_reward}a). Outcome rewards are sparse. Therefore, the effects are less significant on the intermediate steps the agent takes to complete a task. Process rewards \citep{lightman_lets_2023} are provided at each step. Therefore, they are limited to domains where intermediate steps have less variability (e.g. code generation, mathematical calculations). These two types of rewards represent distinct endpoints in the exploration-exploitation landscape \citep{zhang_procvout_2025}. Intermediate reward structures (e.g. staggered reward in Fig. \ref{fig:agent_reward}a) that interpolate between outcome and process signals offer more flexibility for complex, multistage tasks by customizing exploration and exploitation and adapting to the availability of concrete reward in intermediate steps. This reward design mirrors how organizations often provide diverse \textit{performance feedback} \citep{greve_perf_2003} staggered over an extended period from new employee supervision to scheduled evaluations based on concrete deliverables as their competency develops.
\end{exm}
\begin{exm}[High reliability agent engineering] From an organizational view, converting a failure-prone agentic system to a highly reliable one can be seen as a positive organizational change (Fig. \ref{fig:agent_reward}b). The units of organizational behavior are regarded as routines (i.e. conserved action sequences in a fixed order) carried out within the organization \citep{nelson_evolutionary_1982,pentland_routineanalysis_2005}. Managers oversee the completion of routines and provide performance feedback on individual actors \citep{greve_perf_2003}. Reliable execution of routines is seen as a sign of good performance and its inverse as failure. The analysis of failed agent trajectories indicates that they contain highly conserved failure modes \citep{pan_why_2025}. They are, therefore, amenable to solution once we uncover the origins and devise countermeasures. One way to materialize organizational change is to design specific \textit{management interventions} (aka. change intervention) \citep{porras_organization_1991} at the agent level (Fig. \ref{fig:agent_reward}b). The interventions are executed by distinct specialized agents, such as safety guards \citep{xiang_guardagent_2025} and compliance officers, to critique the behavior of worker agents. This helps curtail the emergence of undesirable behaviors due to degraded context sensitivity at long horizon \citep{maharana_evaluating_2024,hong_context_2025} and prevents worker agents from getting lost in their own trajectories \citep{laban_lost_2025}. Alternatively, knowledge about existing high reliability systems can be transferred to (re-)engineering other ones by implementing HR principles into agentic systems. This requires the active participation of every agent within the system to check for potential failures, which comes at the cost of increased capability requirements. Using coding agents as an example, we discuss in Appendix \ref{sec:hro} a potential implementation by updating agent components to strengthen system reliability.

\end{exm}
\begin{tcolorbox}[colback=Gray!20, colframe=Black, left=.2mm, right=.2mm, top=.5mm, bottom=.5mm, sharp corners, boxrule=-1pt, label=rmk3, segmentation style=solid]
\textbf{Remark 3} Machine-compatible agent management should leverage the flexibility of agentic systems and the interplay between internal and external mechanisms to mitigate operational failures and safety threats.
\end{tcolorbox}

\section{Discussion and conclusion}
\label{gen_inst}


Agentic systems operate differently from human organizations (see Table \ref{tab:orgvsags}). Apart from flawed agency, LLM agents are not reliable simulacra of human behavior and values due to their inherent biases \citep{taubenfeld_systematic_2024} and limited grounding. Therefore, an agentic system cannot simply adopt the structure of the human organization it aims to emulate. For application purposes, additional mechanisms are needed to ensure its functioning. Existing approaches to agentic system reliability primarily come from robust machine learning and computer security. They are limited due to the following reasons.

\begin{enumerate}[wide, labelindent=0pt, itemsep=0pt, topsep=1pt]
\item[(1)]Robustness in deep learning is brittle. There are many problems within the existing approaches of certified robustness \citep{wu_toward_2023}. One critical issue is that the current benchmarks are not useful. Because AI agents interact with constantly changing real-world digital infrastructure more often than previous generative AI models, they will encounter more distribution shifts and unexpected failure modes \citep{li_commercial_2025}. Engineering robustness into AI agents operating in the wild will require adaptive, system-specific approaches.
\item[(2)] Agentic systems are more extensive than LLMs. Traditional LLM guardrails are a safety layer situated on top of an existing model to filter responses and block malicious queries \citep{dong_guardrails_2024}. Agentic systems are more extensive than LLMs and they have a much larger attack surface, which is not entirely covered by traditional guardrails. Guardrails are generally built for risk mitigation in human-AI interaction, so they cannot alleviate risks in inter-agent communications.
\end{enumerate}

In conclusion, we uncovered several key aspects to think about the reliability issues facing AI agent engineering through an organizational lens. We find that organizational thinking can provide epistemic foundations for understanding AI agent behavioral traits and collective actions. It is fitting for describing small groups of AI agents with a well-defined power structure customized for different applications. Organization-level approaches complement existing behavior analysis for LLMs and AI agents inspired by social psychology and cognitive science \citep{hagendorff_machine_2023,jackson_ai_2025} and provide concrete paths to designing behavioral interventions for managing and mitigating agent failures that go beyond a security perspective.

Organization science and AI have sustained a relationship since their respective inception in the last century \citep{sycara_multiagent_1998,horling_survey_2004,csaszar_organizations_2022}. The emergence of AI agents powered by LLMs has only reinvigorated the exchange. On the one hand, agents can carry out situated actions for coordinated problem-solving that can be used for simulating, albeit flawed, human interactions within organizations. On the other hand, the accumulated knowledge in organization science over the past decades on studying social systems of humans creating and utilizing technologies, as well as the entrenched issues therein \citep{garicano_why_2016} has assembled valuable knowledge for engineering open-ended, interactive multiagent systems. Aside from informing the study on human-AI collaboration, organization science can also act as a blueprint for the construction of a coherent and systematic theory of AI behaviors to facilitate the safe and effective integration of sociotechnical systems involving humans and AIs \citep{vaccaro_when_2024,riedl_ai_2025}.

\section*{Acknowledgements}
RPX thanks the staff at the Cooperative AI Foundation for inspiring discussions during the inaugural Introduction to Cooperative AI course. We thank D. M. Holtz at Columbia Business School for helpful exchanges on the initial idea.



\newpage
\appendix

\section{Comparison between agentic tool-use architectures}
\label{app:tools}

We compare the three agentic system architectures described in Example \ref{exm:tool_agents} and list the different design aspects they represent in Table \ref{tab:tool_use}. Our goal here is to illustrate the effects on the accountability of the agents from separating the user agent and the tooling system through delegation, here materialized as outsourcing to integrated service providers (i.e. provider bundling) or third-party contractors. The simplification of user agent requirements from single-agent to multiagent tool-use architectures reduces the risk of misalignment behaviors in direct user interaction.
\begin{table*}[htbp!]
\setstretch{1.0}
    \centering
    \begin{tabular}{cccc}
    \toprule
    \multirow{2}{*}{\parbox{2.2cm}{\setstretch{1}\centering \textbf{Agent architecture}}} & \multirow{2}{*}{\parbox{3.3cm}{\setstretch{1}\centering \textbf{User agent requirements}}} & \multirow{2}{*}{\parbox{3cm}{\setstretch{1}\centering \textbf{Tooling agent requirements}}} & \multirow{2}{*}{\parbox{3cm}{\setstretch{1}\centering \textbf{Accountability}}} \\
    & & & \\
    \midrule
    \multirow{5}{*}{\parbox{2.2cm}{\setstretch{1}\centering Single-agent\\tool use}} & \multirow{5}{*}{\parbox{3.8cm}{\setstretch{1}\centering Single agent can select tools (e.g. full selection), understand how to use them for different tasks}} & \multirow{5}{*}{\parbox{3.2cm}{\setstretch{1}\centering None (user agent interacts directly with tools at all times)}} & \multirow{5}{*}{\parbox{3.6cm}{\setstretch{1}\centering User agent accounts for all actions}} \\
    & & & \\
    & & & \\
    & & & \\
    & & & \\
    \hline
    \multirow{6}{*}{\parbox{2.2cm}{\setstretch{1}\centering MAS with provider-bundled agents}} & \multirow{6}{*}{\parbox{3.3cm}{\setstretch{1}\centering Single agent can select tool provider (e.g. partial selection) and communicate tasks with provider agents}} & \multirow{6}{*}{\parbox{3.2cm}{\setstretch{1}\centering Provider agents can understand tool functionalities, usage cases, and distinctions between tools}} & \multirow{6}{*}{\parbox{5.6cm}{\setstretch{1}\centering Use agent and provider agents separately accounts for tool-use failures; User agent accounts for provider selection failure; Each provider is responsible for its own bundled tools}} \\
    & & & \\
    & & & \\
    & & & \\
    & & & \\
    & & & \\
    \hline
    \multirow{7}{*}{\parbox{2.2cm}{\setstretch{1}\centering MAS with supportive\\tooling agents}} & \multirow{7}{*}{\parbox{3.3cm}{\setstretch{1}\centering Single agent can communicate tasks with tooling agents}} & \multirow{7}{*}{\parbox{3.6cm}{\setstretch{1}\centering Tooling agents can work in a team with distinct roles for each member (e.g. orchestrator, planner, retriever, verifier)}} & \multirow{7}{*}{\parbox{5.6cm}{\setstretch{1}\centering Tooling system generally accounts for tool-use failures; Each tooling agent (either from the same or distinct providers) can be individually assessed to account for failure in its role-defined stage of execution}} \\
    & & & \\
    & & & \\
    & & & \\
    & & & \\
    & & & \\
    & & & \\
    \bottomrule
    \end{tabular}
    \vspace{0.5em}
    \caption{Tool-use agentic systems with distinct delegational structures and their potential issues}
    \label{tab:tool_use}
\end{table*}
\begin{table*}[htbp!]
\setstretch{1.0}
    \centering
    \begin{tabular}{ccc}
    \toprule
    \multirow{2}{*}{\parbox{4.2cm}{\setstretch{1}\centering \textbf{High reliability principle}}} & \multirow{2}{*}{\parbox{3.3cm}{\setstretch{1}\centering \textbf{Type of change}}} & \multirow{2}{*}{\parbox{6cm}{\setstretch{1}\centering \textbf{Implementation in coding agents}}} \\
    & & \\
    \midrule
    \multirow{4}{*}{\parbox{4.2cm}{\setstretch{1}\centering (i) Preoccupation with failure}} & \multirow{4}{*}{\parbox{3.3cm}{\setstretch{1}\centering Reward change}} & \multirow{4}{*}{\parbox{7.5cm}{\setstretch{1}\centering Allocate the highest reward to the successful execution of the entire generated code and reduce or skip stepwise reward}} \\
    & & \\
    & & \\
    & & \\
    \hline
    \multirow{4}{*}{\parbox{4.2cm}{\setstretch{1}\centering (ii) Reluctance to simplify}} & \multirow{4}{*}{\parbox{3.3cm}{\setstretch{1}\centering Structure change}} & \multirow{4}{*}{\parbox{7.5cm}{\setstretch{1}\centering Warrant an operation only through execution, such as accessing a terminal for coding agents \citep{wang_executable_2024} before committing the change}} \\
    & & \\
    & & \\
    & & \\
    \hline
    \multirow{4}{*}{\parbox{4.2cm}{\setstretch{1}\centering (iii) Sensitivity to operations}} & \multirow{4}{*}{\parbox{3.3cm}{\setstretch{1}\centering Interaction change}} & \multirow{4}{*}{\parbox{7.5cm}{\setstretch{1}\centering Examine critical parts of the generated code by multiple agents, such as through debate \citep{li_swedebate_2025}, and come to a quality assessment afterwards}} \\
    & & \\
    & & \\
    & & \\
    \hline
    \multirow{4}{*}{\parbox{4.2cm}{\setstretch{1}\centering (iv) Commitment to resilience}} & \multirow{4}{*}{\parbox{3.3cm}{\setstretch{1}\centering Structure change}} & \multirow{4}{*}{\parbox{7.5cm}{\setstretch{1}\centering Allocate a separate agent specifically designed for bug detection and correction in the generated code. This can be combined with (ii)}} \\
    & & \\
    & & \\
    & & \\
    \hline
    \multirow{5}{*}{\parbox{4.2cm}{\setstretch{1}\centering (v) Deference to expertise}} & \multirow{5}{*}{\parbox{3.3cm}{\setstretch{1}\centering Interaction change}} & \multirow{5}{*}{\parbox{7.5cm}{\setstretch{1}\centering Allow subagents to disobey a head agent through negotiation \citep{fatima_principles_2014,bianchi_how_2024} especially when the task is unsafe or that the subagent lack the tools to carry it out properly}} \\
    & & \\
    & & \\
    & & \\
    & & \\
    \bottomrule
    \end{tabular}
    \vspace{0.5em}
    \caption{Potential realization of high reliability principles in coding agents through organizational changes.}
    \label{tab:hr_agents}
\end{table*}

\section{Maintenance of agentic systems}
The following are characteristic types of agentic systems powered by LLMs with different levels of complexity. We provide a practical example for each type and consider the potential cost (monitoring and adaptation) involved in their respective maintenance procedure.

\textbf{(AI-augmented) workflows} are structured action sequences determined before execution. Some steps within the sequence are executed by an AI model\footnote{\url{https://www.anthropic.com/engineering/building-effective-agents}}
\begin{enumerate}[itemsep=0pt, topsep=1pt]
    \item[] \textbf{Example} An Excel sheet AI agent that converts unstructured text into table entries.
    \item[] \textbf{Monitoring cost} is low because these workflows involve highly conserved actions (e.g. routines) and the system does not exhibit a pronounced level of intelligence. The task execution is highly repetitive and contains very few edge cases.
    \item[] \textbf{Adaptation cost} is high because most of the component in the workflow needs to change to accommodate a new task.
\end{enumerate}

\textbf{AI assistants} are interactive user interfaces powered by LLMs or large AI models with language output. They are also called \textbf{(AI) copilots}.
\begin{enumerate}[itemsep=0pt, topsep=1pt]
    \item[] \textbf{Example} An airline customer service AI agent that can help users book flights, reschedule or cancel flight bookings, and check current airline and flight information from the airline's internal database.
    \item[] \textbf{Monitoring cost} is moderate since the agent operates with an increased amount of potential edge cases compared to simple workflows.
    \item[] \textbf{Adaptation cost} is moderate since one can restrict the agent by forbidding the agent from executing certain tasks (e.g. by directing it to humans) or letting it learn new ones.
\end{enumerate}

\textbf{Domain-specialized AI agents} are individual agents operating with a level of autonomy on a specific task and can use capabilities such as planning, reasoning, and tool use to complete the task through sequential interaction with the environment.
\begin{enumerate}[itemsep=0pt, topsep=1pt]
    \item[] \textbf{Example} An AI study partner that can interact with users and the internet to seek relevant course and practice materials. It can also generate course plans and help the user unpack difficult materials with easy explanations and lay summaries.
    \item[] \textbf{Monitoring cost} is high because of the level of autonomy prescribed to the agent. 
    \item[] \textbf{Adaptation cost} is moderate because the agent already has access to relevant tools and has relevant capabilities to carry out information-seeking tasks to correct its own output. Adaptation only needs to consider abrupt changes in the deployment environment (e.g. updates of an external database login page) and slow drifts in the user preference (e.g. user becomes more interested in a discipline the agent is not very knowledgeable in at the start).
\end{enumerate}

\textbf{General-purpose agentic systems} are capable of completing a wide range of tasks through the creation of and interaction between multiple (sub)agents. The system can be constructed using a single (head) agent, which then spawns additional worker agents during task execution. Alternatively, it can be initialized with a team of agents with no specific role differentiation, but they can take on specific roles based on the task specification and resources provided alongside it by the human users \citep{lai_collective_2024}. General-purpose MAS is considered close to a fully autonomous agent \citep{mitchell_fully_2025}.
\begin{enumerate}[itemsep=0pt, topsep=1pt]
    \item[] \textbf{Example} An agentic taskforce which helps the user or employer with a great number of complex tasks (e.g. grow a healthy habit, consult for (mental) health problems, book tickets for travels and entertainment).
    \item[] \textbf{Monitoring cost} is high because of the great number of capabilities the agentic system has.
    \item[] \textbf{Adaptation cost} is low because the agentic system can learn to adapt through self-critiquing and gaining knowledge from digital online resources with minimal human intervention.
\end{enumerate}

\section{High reliability coding agents}
\label{sec:hro}
We use an agentic system for coding (i.e. code generation) as an example for discussing the implementation of an organizational change to engineer high reliability agentic systems. Coding tasks exhibit great varieties and lengths \citep{jiang_codeagents_2025}, their failure modes can vary. We consider the organizational change from three different aspects: structure, reward, and interaction. A potential implementation is discussed in Table \ref{tab:hr_agents}. Although separate parts may have preliminary implementation, to fully realize the HR principles requires integrating all relevant changes in a single system.


\bibliography{example_paper}
\bibliographystyle{icml2025}

\end{document}